# Integration of iron in natural and synthetic Al-pyrophyllites: an infrared spectroscopic study.

**Sébastien Lantenois** [1,2] *

**Jean-Michel Bény**[1]

**Fabrice Muller**[1]

**Rémi Champallier**[1]

[1] Institut des Sciences de la Terre d'Orléans (ISTO), CNRS – Université d'Orléans, 1A rue de la Férollerie, 45071 Orléans Cedex 2, France.

[2] Laboratoire des Agrégats Moléculaires et Matériaux Inorganiques (LAMMI), CNRS – Université Montpellier 2, Bât 15 – Case courrier 015, Place Eugène Bataillon, 34095 Montpellier CEDEX 5 – France.

* Author to whom correspondence should be addressed.

e-mail: sebastien.lantenois@univ-orleans.fr

ABSTRACT

Numerous studies focus on the relationships between chemical composition and OHband positions in the infrared (IR) spectra of micaceous minerals. These studies are based on the coexistence, in dioctahedral micas or smectites, of several cationic pairs around the hydroxyl group which each produce a characteristic band in the IR spectrum. The aim of this work is to obtain the wavenumber values of the IR OH vibration bands of the (Al-$Fe^{3+}$)-OH and ($Fe^{3+}$-$Fe^{3+}$)-OH local cationic environments of 'pyrophyllite type' in order to prove, disprove or modify a model of dioctahedral phyllosilicate OH-stretching band decomposition. Natural samples are characterized by powder X-ray diffraction (XRD), Fourier transform infrared (FTIR) and Raman spectroscopies and electron microprobe; the hydrothermal synthesis products are also analysed by powder XRD and FTIR after inductively coupled plasma measurements to obtain the chemical compositions of nascent gel phases. Natural samples contain some impurities which were eliminated after acid treatment; nevertheless, a small Fe content is found in the pyrophyllite structure. The amount of Fe which is incorporated within the pyrophyllite structure is much more important for the synthetic samples than for the natural ones. The IR OH bands were clearly observed in both natural and synthetic pyrophyllites and assigned to hydroxides bonded to (Al-Al), (Al-Fe) and (Fe-Fe) cationic pairs. During this study, three samples were analysed by DTG to check the cis- or trans-vacant character of the layers and to determine the influence of this structural character on the OH-stretching band position in IR spectroscopy.

**KEYWORDS:** Pyrophyllite, hydrothermal syntheses, FTIR spectroscopy, iron defaults, cis-trans-vacant.

# INTRODUCTION

Dioctahedral phyllosilicates have a wide spectrum of chemical composition [1] and their cations distribute with different degrees of order − disorder in the tetrahedra and octahedra of the sheet structure. Infrared spectroscopy is an efficient tool for the determination of local cationic environments as well as fine structural features. The main relationships between chemical composition and OH bands positions in infrared spectra of various dioctahedral materials have been established for celadonites and glauconites [2-5], micas [6-8] or smectites [9-14]. With the computer science development it is really possible to decompose the broad infrared band of O-H vibrations of phyllosilicates. The first model of (cation-cation)-OH vibrations corresponding to different wavenumber bands in the OH stretching range was proposed by Besson *et al*. [6-7] for dioctahedral *trans*-vacant mica and more recently adapted to dioctahedral smectites by Zwiagina *et al*. [13]. These models were based on the coexistence in dioctahedral micas or smectites of (cation-cation)-OH bands linked to an interlayer cation (mica-type bands) and three bands not linked to an interlayer cation (pyrophyllite type bands) [6]; Al-Al-OH, Al-$Fe^{3+}$-OH and $Fe^{3+}$-$Fe^{3+}$-OH bands. The Al-Al-OH stretching band position in pyrophyllite ($Si_4Al_2O_{10}OH_2$) is clearly identified [2, 15-17] in comparison with the Fe-pyrophyllite OH bands ones. The scarce occurrence of natural ferripyrophyllite (iron rich equivalent of pyrophyllite, described by Choukhrov *et al*. [18]) and the very low content of iron substitution in natural Al-pyrophyllite structure explain the lack of Fe-pyrophyllite infrared studies.

The identification of these OH pyrophyllite vibrations are necessary to approve disapprove or modify the model of dioctahedral phyllosilicate OH stretching decomposition. The aim of this study is to identify the infrared OH vibration bands in Fe-pyrophyllite structure. Two complementary approaches were carry-out: the first one consisting to the study of natural Fe-pyrophyllites and the second one using hydrothermal synthetic samples. These approaches have been realized with fast Fourier transform infrared (FTIR) spectroscopy coupled with chemical analyses, transmission electron microscopy (TEM), powders X-ray diffraction (XRD) and Raman spectroscopy.

MATERIALS AND METHODS

*Materials*

A variety of natural and synthetic samples were used in this study. The natural samples are described in Table 1. Synthetic pyrophyllites were crystallized from gels under hydrothermal conditions. Si-Al gels were prepared following the usual method adapted from Hamilton and Henderson [19]. Tetraethylorthosilicate (TEOS), $Al(NO_3)_3 \cdot 9H_2O$, $HNO_3$, $NH_4OH$ and ethanol were used as reagents with a 99% minimum grade purity. After dissolution of Al-nitrate in nitric acid, TEOS and ethanol were added. A precipitate was obtained by neutralizing the resulting solution at pH ~6 with addition of $NH_4OH$. This precipitate was dried at 80°C for 24 hours, ground in an agate mortar and heated up to 400°C to remove nitrates and to obtain a gel composed essentially of Al and Si oxides. Si-Al-Fe gels were prepared using sodium metasilicate ($SiO_2 \cdot Na_2O \cdot 5H_2O$), $AlCl_3 \cdot xH_2O$, $FeCl_3 \cdot 6H_2O$ and HCl [20] according to the reaction: $4(SiO_2 \cdot Na_2O) + x(AlCl_3) + (2 - x)(FeCl_3) + 2HCl \rightarrow Si_4Al_xFe^{3+}_{2-x}O_{11} + 8NaCl + H_2O$. Two x values are used : 0.40 and 1.75. After precipitation, the solid phase was washed and centrifuged to remove sodium chloride (five washing). Precipitates were dehydrated by drying at 60°C during 24 hours and ground in an agate mortar to obtain a gel composed essentially of Si, Fe and Al oxides [21]. The samples were synthesized in an internally heated pressure vessel. 100 mg of gel were mixed to 100 mg of water in a gold tube. The tube was then sealed and heated under argon pressure. The experimental conditions and the chemical composition of the beginning gel phases (determined by ICP) are reported in table 2. After cooling of the vessels, the solid products were extracted from the reaction tubes, dried at 80°C overnight, and ground before structural and chemical characterizations.

Some natural pyrophyllites containing additional phases were acid treated to remove these phases. Treated samples were prepared following an acid attack method adapted from Perez-Rodriduiez *et al.* [22] and Maqueda *et al* [23]; known to destabilize phyllosilicate phases except pyrophyllite : A mixture of $HNO_3$, HF and $HClO_4$ was prepared. 100 mg of finely ground samples were placed in a Nalgène® reactor, mixed with 20 ml of deionized water, 2.5 ml of $HNO_3$ (68% weight), 2.5 ml of $HClO_4$ (60% weight) and 5 ml of HF (40% weight) and heated at 80°C during three hours. The residual solid part was filtered, washed with 100 ml of deionized water, dried by heating at 80°C during one hour and then ground before characterization.

*Methods:*

Natural and synthetic pyrophyllites were characterized by X-ray powders diffraction. XRD patterns were recorded in transmission geometry using CoKα radiation (35 mA, 35 kV). The use of an INEL CPS 120 curved position sensitive detector allowed simultaneously recording the diffracted intensity over a 4-50°2θ range with a step size of 0.03°. The non-linearity of the detector was corrected [24]. A 0.5 mm diameter Lindemann glass tube was used to hold the sample powder.

FTIR spectra of natural pyrophyllites were recorded over the 650–4000 $cm^{-1}$ range with a 2 $cm^{-1}$ resolution using a Nicolet Magna 760 IR Fourier transform spectrometer equipped with a Globar SiC source and a DTGS detector. The spectrometer was purged with dry air to remove most of atmospheric $H_2O$ and $CO_2$ during data collection. Two recording types were investigated. For natural samples the classical pellet method was used : the sample was finely ground in an agate mortar, and 0.5 mg of the resulting powder was mixed with 149.5 mg of KBr previously dried at 120°C for 24 hours. The mixture was homogenized and pressed in an evacuable die to prepare a 12 mm diameter pellet. For synthetic samples, infrared spectra were recorded using the same Fourier transform spectrometer coupled with a Nicolet Nic-Plan microscope to optimize IR signal and limit the amount of used sample. The powder was spread over the NaCl window of the microscope stage. The analysed sample area was a 100 μm diameter circle chosen under the microscope 15 X Cassegrainian objective. The operating conditions were 200 scans, 2 $cm^{-1}$ resolution over the 650 – 4000 $cm^{-1}$ range without ambient $H_2O$ correction.

The chemical composition of the pyrophyllite samples were realized using a CAMECA SX50 electron microprobe (EMPA), operating at 15 kV accelerating voltage. Analyzed samples were grounded in an agate mortar and pressed in an evacuable die to obtain a flat pellet necessary to obtain best electron microprobe analyses. A part of these pellets was fixed on a glass slide and then silver metalized befor analyze.

Raman spectra were obtained using a Dilor XY 800 confocal micro Raman spectrometer equipped with a Wright Model charge-coupled detector (CCD). The excitation source was the 514 nm green line of a Coherent Innova Model 90-5 $Ar^+$ laser. The beam was focused onto samples powder placed in a glass slide using a 100X objective. Measurements were realized with a laser power of 100 mW and an integration time of 300s.

RESULTS

*Defaults in natural pyrophyllites*

The infrared spectra (Figure 1) in the OH stretching zone (3600-3700 cm$^{-1}$) of some pyrophyllite samples (Table 1) indicate the presence of structural defaults. The Al-Al-OH stretching band of pyrophyllite is classically fixed near 3674 cm$^{-1}$ ([2], Figure 1a). Associated to this intense vibration band, other ones at 3620, 3630, 3645, 3655, 3668 and 3698 cm$^{-1}$ were detected (Figure 1b-e). Some of these vibration bands are associated with additional phyllosilicate phases. XRD patterns (Figure 2) also show the presence of associated mineral phases. Quartz was identified in Rob49 sample (Figure 2b) and Guad sample (Figure 2c). Some other phyllosilicate minerals were observed such as kaolinite in Sued sample (Figure 2e) and muscovite in Rob49 sample (Figure 2b) and Sued sample (Figure 2e). The associated phases detected by XRD are listed in Table 1.

These associated phases were eliminated with an acid treatment which allows kaolinite and mica phases dissolving without destabilization of the pyrophyllite phase. Chemical analyze results using electron microprobe technique of the acid treated pyrophyllite samples are given in table 3. The disappearance of the OH stretching bands corresponding to the associated phases is confirmed by FTIR recording of the acid treated samples (Figure 3). The large band characteristic of muscovite near 3630 cm$^{-1}$ in Rob49 sample (Figure 1b) is missing (Figure 3b). In Sued sample, the bands associated to kaolinite at 3620, 3655 and 3698 cm$^{-1}$ (Figure 1e) are not observed after acid treatment (Figure 3e). Nevertheless, some additional vibration bands were always present : a shoulder near 3668 cm$^{-1}$ in Guad sample (Figure 1c and 3c) and a small band at 3645 cm$^{-1}$ in SN and Sued samples (Figure 1d,e and 3d,e). Moreover the vibration band at 3645 cm$^{-1}$ was also identified in three other samples presented in this study (Figure 3f,g,h) and was always correlated to the iron detection by electronic microprobe analyze (Table 3).

The OH bending zone (800-1000 $cm^{-1}$) exhibit a band characteristic of pyrophyllite at 950 $cm^{-1}$ [2] in all samples (Figure 4a-h). Three other bands at 816, 836 and 855 $cm^{-1}$ were always present and generally observed in pyrophyllite [2]. A very small band at $\cong$ 907 $cm^{-1}$ could also be observed in some pyrophyllites samples (Figure 4d-h) and was systematically correlated to the band at 3645 $cm^{-1}$ in the OH stretching zone (Figure 3d-h) and to the presence of iron (Table 3).

*Integration of iron in synthetic pyrophyllites*

An experimental run of hydrothermal syntheses was realized in order to synthesize pyrophyllite with different iron content. Experimental conditions were reported in Table 2 and in accordance with the stability domain of Al-pyrophyllite identified by Eberl [25] and established by Kloprogge *et al*. [15]. XRD patterns of synthetic samples were presented in Figure 5. Only pyrophyllite reflections were identified (Figure 5a) for the composition (PAl-1). The second Al-pyrophyllite (PAl-2), synthesized at different temperature/pressure conditions (Table 2), shown the same XRD pattern as PAl-1 sample. After hydrothermal treatment, the two Al-Fe samples (PFe-1 and PFe-2) were composed of a mixture between hematite ($\alpha$-$Fe_2O_3$), opal ($SiO_2$, $xH_2O$) and pyrophyllite (Figure 5c,d). The next synthetic compound, without Al content (PFe-3), was composed of hematite and opal (Figure 5e).

The Infrared spectra of the synthetic pyrophyllite samples are presented in Figure 6 (OH stretching zone) and Figure 7 (OH bending zone). The presence of an OH stretching vibration band at 3674 $cm^{-1}$ was observed in Al-pyrophyllite (PAl-1 and PAl-2) but was associated to another band at 3668 $cm^{-1}$ (Figure 6a,b). The 3668 $cm^{-1}$ band was previously observed in the Guadeloup natural pyrophyllite (Figure 1c, 3c). The intensity ratio between these two bands was different in PAl-1 sample and in PAl-2 one. In PFe-1 and PFe-2 samples, theses bands were still present but in addition two other ones at 3645 cm$^{-1}$, like in natural pyrophyllite with iron content (Table 3, Figure 3d-h) and at 3620 cm$^{-1}$ were detected (Figure 6c, d). In the OH bending zone, a band near 950 cm$^{-1}$ was observed in all synthetic samples (Figure 7a-d) as in natural ones (Figure 4). In PAl-1 and PAl-2 samples, the three bands at 816, 836 and 855 cm$^{-1}$ characteristic of pyrophyllite [2] are presents but two other bands were observed at 919 and 882 cm$^{-1}$ (Figure 7a,b). In PFe-1 and PFe-2 samples, two bands at 903 and 875 cm$^{-1}$ were detected and the presence of Si-O vibration band of Opal near 800 cm$^{-1}$ was also observed (Figure 7c,d).

## DISCUSSION

*The AlAlOH vibration bands:*

*Identification of bands in the OH stretching zone:*

Generally, natural pyrophyllites are characterized with an infrared absorption peak at $\cong$ 3675 cm$^{-1}$ (Al-Al-OH vibration band, [2]) and a dehydroxylation maximum temperature between 550 and 680°C [26-28, 42].

The infrared characteristic band was always observed for all the natural and synthetic pyrophyllites presented in this study (Figure 1, 3 and 6). However, another hydroxyls environment was detected with a band close to 3668 cm$^{-1}$ identified as a shoulder in one natural sample (Guad sample, Figure 3c) and as a peak in some synthetic ones (Figure 6). Because the synthetic samples PAl-1 and PAl-2 containing only aluminum and silicon (Table

2) and only pyrophyllite phase (Figure 5a,b), the band at 3668 cm$^{-1}$ can only be associated to an Al-Al-OH vibration of pyrophyllite structure.

The presence of two structural hydroxyls alumina-environments was also observed in the thermal curves. The DTG curves of the three samples (Guad, PAl-1 and PAl-2), presented in figure 8 showed two dehydroxylation peaks between 400 and 850°C. The more intense peak was observed near 560°C and the second peak near 780°C. The existence of two different dehydroxylation temperatures is currently observed in some dioctahedral 2:1 phyllosilicates [30-32] and is correlated to the existence of *cis*- or *trans*-vacant layers [32,33]. Indeed, the layers of dioctahedral 2:1 phyllosilicates can differ in the distribution of the octahedral cations over the *cis*- and *trans*-vacant sites. In dioctahedral phyllosilicates, the octahedral sheet is composed of two cations and a vacancy around the OH group in the octahedral layers. If the two cations were placed in $M_2$ and $M_{2'}$ sites (Figure 9) the structure is identified as a *trans*-vacant structure. On the contrary, if one cation is placed in $M_1$ site and the second in $M_2$ (or $M_{2'}$) site, the structure is identified as a *cis*-vacant structure. Drits *et al.* (1995) assumed that thermal energy needed for a proton to jump to the nearest OH group to form a water molecule strongly depends on the distance between the nearest OH groups. Therefore *cis*-vacant structure requires higher dehydroxylation energy than *trans*-vacant one because of the longer OH-OH distance [33]. This affirmation is confirmed in the literature in which *trans*-vacant dioctahedral phyllosilicates dehydroxylate between 550 and 680°C and *cis*-vacant ones dehydroxylate between 700 and 850°C [26, 30-40].

Although, pyrophyllite minerals are classically identified as *trans*-vacant phyllosilicates [33,41] with a dehydroxylation maximum between 550 and 680°C [26-28,42], Wang and Zang [43, 44] had identified three kinds of pyrophyllites populations. The first one has a large DTG dehydroxylation peak with a maximum temperature near 650°C and is identified as *trans*-vacant structure. The second one has a large DTG dehydroxylation peak with a maximum temperature near 880°C and is identified as *cis*-vacant structure. The third one presents two maxima at 650 and 880°C which correspond to a *cis*- and *trans*- mixture of pyrophyllite phases or an interstratification between *cis*- and *trans*-layers. For the three samples Guad, PAl-1 and PAl-2, the two structural OH losses near 550 and 750°C could be interpreted as the presence of *trans*- and *cis*-vacant layer respectively. So, it can be considered that the presence of these two OH stretching bands will be correlated to the existence of a structure partially *cis*- and *trans*-vacant. In this case, the band at 3674 cm$^{-1}$ will be characteristic of *trans*-vacant Al-Al-OH bands and the band at 3668 cm$^{-1}$ will be characteristic to *cis*-vacant Al-Al-OH.

*Correlation in the OH bending zone:*

In natural pyrophyllite (with *trans*-vacant structure [33, 41]), the Al-Al-OH band in the OH bending zone is classically identified at about 950 cm$^{-1}$ [45]. This band was observed in all the pyrophyllite samples presented in this study (Figure 4,7).

For Al-synthetic samples (PAl-1 and PAl-2) two other bands were identified at 882 and 919 cm$^{-1}$ (Figure 7a,b). The band near 880 cm$^{-1}$ is not included in the bending vibration range of Al-Al-OH (between 900 and 955 cm$^{-1}$) but is observed in a lot of synthetic dioctahedral 2:1 phyllosilicates (beidellite, pyrophyllite and montmorillonites, Klopprogge *et al*. [15, 46], Lantenois *et al*. [47]). Because Russell *et al*. [45] have observed this band in a synthetic deuterated pyrophyllite sample, it can not be attributed to an OH vibration band.

The band about 919 cm$^{-1}$ is included in the bending range. Because the band at 950 cm$^{-1}$ is classically observed in *trans*-vacant pyrophyllites, we attribute the 919 cm$^{-1}$ band to the presence of *cis*-vacant layers of pyrophyllite. This interpretation is in accordance with the Al-Al-OH bending band positions of different *cis*- and *trans*-vacant dioctahedral 2:1 phyllosilicates. Among dioctahedral smectites, most montmorillonites, composed by *cis*- vacant 2:1 layers [48] have the Al-Al-OH bending range between 910 and 920 cm$^{-1}$ [14, 47, 49, 51]. For most beidellites, composed by *trans*-vacant layers [48]. this range is 930 - 940 cm$^{-1}$ [2, 14, 52].

*The OH vibration bands linked to the iron incorporation:*

*Identification of bands in the OH stretching zone:*

In natural and synthetic studied samples with iron content (Table 3), a band at 3645 cm$^{-1}$ was always observed (Figure 3). This band is interpreted by Farmer [1964] as the symmetric OH vibration (the two hydroxyls group vibrations are in phase) coupled to the antisymmetric one at 3675 cm$^{-1}$ (the two hydroxyls group vibrations are in phase opposition). In Raman spectroscopy the symmetric and antisymmetric bands have generally an intensity ratio inversed as in infrared spectroscopy. The Raman spectrum of VAN pyrophyllite sample (Figure 10) shows that the symmetric and antisymmetric bands have the same wavenumber and are in the same intensity ratio than in infrared spectrum (Figure 3g). This result permits us to don't pay attention to this interpretation. An alternative assignment is proposed by Farmer [1964] and Besson *et al.* [7]. These authors supposed that the band at 3645 cm$^{-1}$ could be attributed to an Al-Fe$^{3+}$-OH band. Moreover, Besson *et al.* [7] had developed a model which take into account the Al-Fe$^{3+}$-OH band at 3652 cm$^{-1}$ and a Fe$^{3+}$-Fe$^{3+}$-OH band at 3630 cm$^{-1}$ correlated to an increasing of the iron content.

In synthetic PFe-1 and PFe-2 samples a band near 3620 cm$^{-1}$ was detected (Figure 6c, d). In these samples only opal and hematite associated phases are detected with XRD (Figure 5C, d). The only OH bearing phase is a iron content pyrophyllite and then the 3620 cm$^{-1}$ band is associated to a hydroxyls with iron-environment. In the used hydrothermal conditions, iron previously introduced in ferric form can not change during the synthesis [20]. Then, in accordance with Farmer [1964] and Besson *et al.* [7], if the 3645 cm$^{-1}$ is assigned to the Al-Fe$^{3+}$-OH band, the 3620 cm$^{-1}$ band is necessary associated to Fe$^{3+}$-Fe$^{3+}$-OH. This interpretation is confirmed with chemical analyses which shows for synthetic samples (Table 2) an increase of iron content in comparison with the natural ones (Table 3).

*Comparison with the iron analogue of pyrophyllite: the ferripyrophyllite*

Ferripyrophyllite is the ferric analogue of pyrophyllite with an ideal formula Fe$^{3+}_2$Si$_4$O$_{10}$(OH)$_2$. This mineral has a very low occurrence and was only described and characterized in natural state by Chukhrov e*t al.* [18] and synthesized by Grauby [53]. We tried to synthesize ferripyrophyllite in the conditions fixed by Grauby [53] (data not shown) and in hydrothermal conditions used for synthesize our pyrophyllites : (430°C, 0,5 kbar, 29 days) and (475°C, 2 kbar, 15 days). Ferripyrophyllite never appeared in these conditions.

The infrared study of these minerals shows two bands in the OH stretching zone. The first one has a strong intensity and is fixed near 3590 cm$^{-1}$ and the second one has a lower intensity and appeared near 3630 cm$^{-1}$ [18]. These two bands can be attributed to Fe$^{3+}$-Fe$^{3+}$-OH band and an Al-Fe$^{3+}$-OH one respectively, in accordance with the chemical composition of this mineral established by Coey *et al.* [54]. In synthetic sample, only the Fe$^{3+}$-Fe$^{3+}$-OH band was identified by Grauby [53] near 3595 cm$^{-1}$. These data show shifts between Al$^{3+}$-Fe$^{3+}$-OH (15cm$^{-1}$) and Fe$^{3+}$-Fe$^{3+}$-OH (25cm$^{-1}$) band positions in ferripyrophyllites [18, 53] and in our synthetic samples.

These shifts can be explained by the difference in chemical composition. This phenomenon has already been observed in other phyllosilicates. For example, Wilkins and Ito [55] observed shifts lower than 5 cm$^{-1}$ between the stretching vibration bands of Mg-Ni talc and propose to correlate the OH vibration band positions to the ionic radius of octahedral cations. These shift was smaller than those obtained in this study, but the difference of ionic radii between magnesium and nickel is very weak (inferior to 0.03Å, [56]) in comparison with the difference between aluminum and ferric iron (0.11Å [56]).

*Identification of bands in the OH bending zone:*

In natural pyrophyllite samples who containing structural iron, we observed systematically the presence of a band at 907 cm$^{-1}$ (Figure 4). This band was identified as an $Al^{3+}$-$Fe^{3+}$-OH bending band. This band was shifted at 903 cm$^{-1}$ in synthetic pyrophyllite samples (Figure 7c,d). $Fe^{3+}$-$Fe^{3+}$-OH vibration band has been identified by Chukhrov *et al*. [18] and Grauby [53] between 800 and 805 cm$^{-1}$. In this range, the presence of Si-O vibrations bands and the three bands at 856, 837 and 815 cm$^{-1}$ currently observed in pyrophyllite and not assigned to OH bending vibration [45] hindered the vibration band. In addition, in your sample, a synthetic broad band make impossible the identification of this $Fe^{3+}$-$Fe^{3+}$-OH band (Figure 7c,d).

CONCLUSION

ACKNOWLEDGMENTS

P. Baillif, O. Rouer (ISTO-Orléans), A. Pineau (CRMD-Orléans), L.-C. de Menorval (LAMMI-Montpellier) and C. Reibel (LPMC-Montpellier) are thanked for their assistance.

FIGURE CAPTIONS

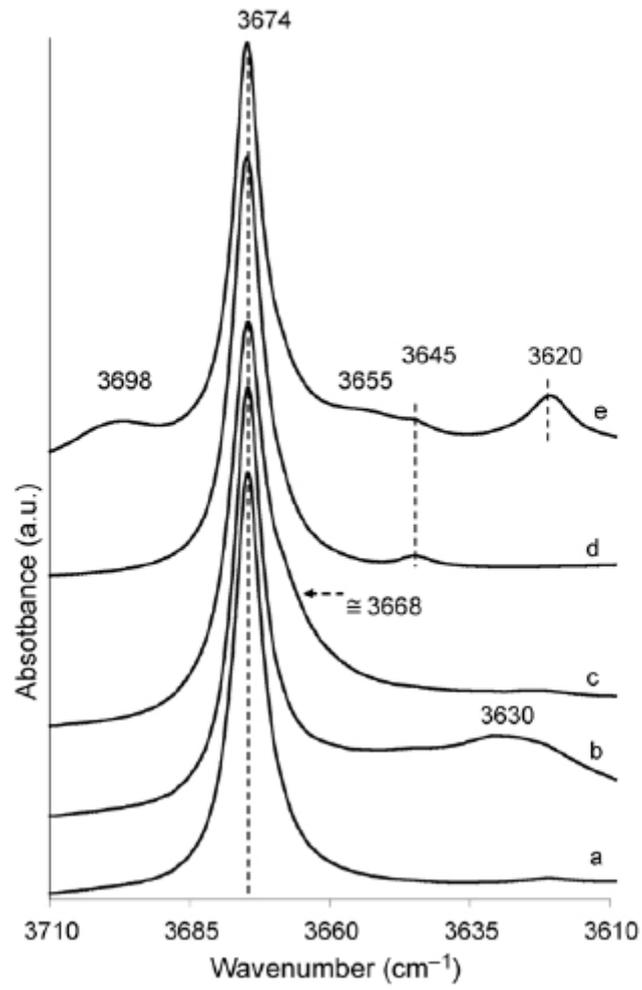

**Figure 1:** Infrared spectra of natural pyrophyllites in the OH stretching zone. (a) Rob 48, (b) Sued, (c) Guad, (d) SN, (e) Rob 49.

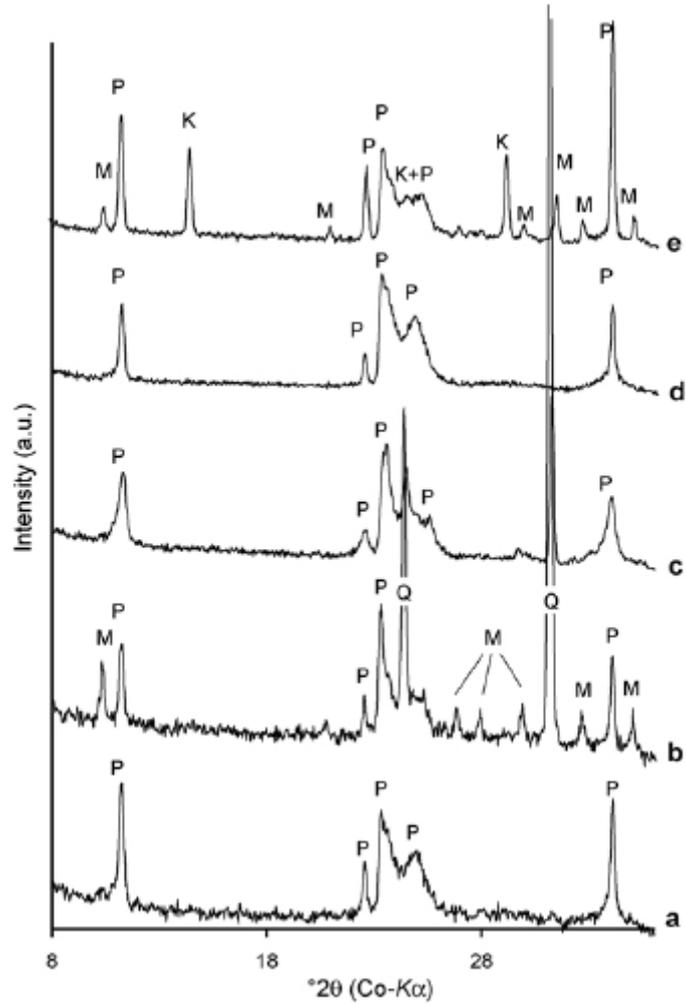

**Figure 2:** XRD patterns of natural pyrophyllites. (a) Rob 48, (b) Sued, (c) Guad, (d) SN, (e) Rob 49. Pyrophyllite reflections are labeled P. Quartz, muscovite and kaolinite impurities are labeled Q, M and K respectively.

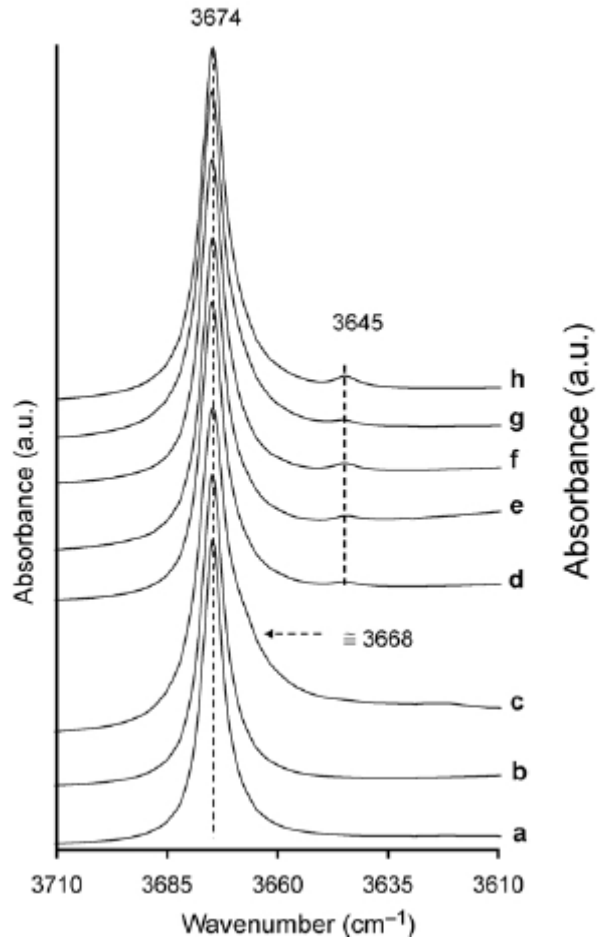

**Figure 3:** Infrared spectra of pyrophyllites in the OH stretching zone after acid treatment (noted -T) if necessary. (a) Rob 48-T, (b) Rob 49-T, (c) Guad-T, (d) SN-T, (e) Sued-T, (f) Bin (g) Nep, (h) Van. For the three last samples, it wasn't necessary to perform acid treatment because phyllosilicate phases were not present in these samples.

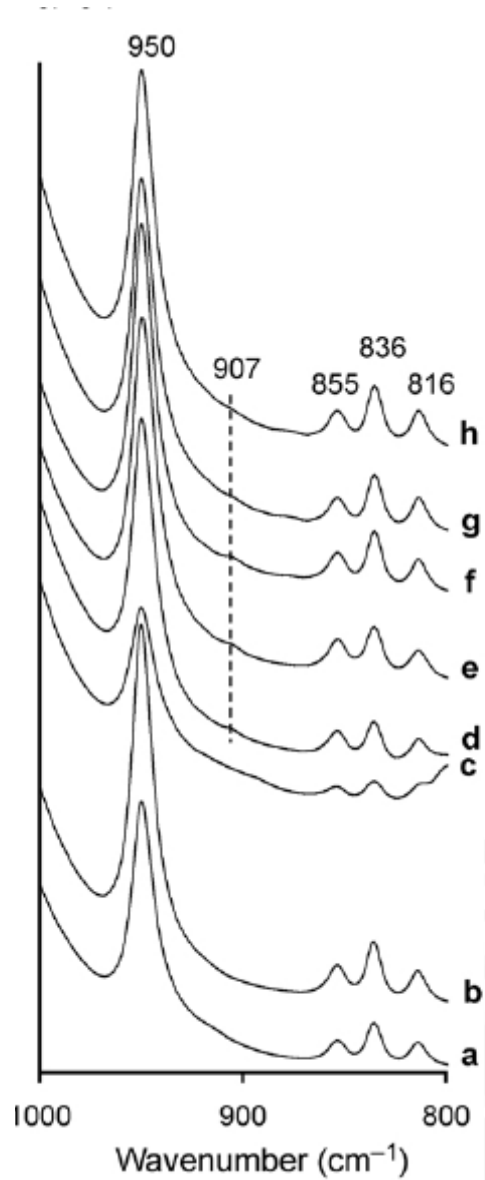

**Figure 4:** Infrared spectra of pyrophyllites in the OH bending zone after acid treatment (noted -T) if necessary. (a) Rob 48-T, (b) Rob 49-T, (c) Guad-T, (d) SN-T, (e) Sued-T, (f) Bin (g) Nep, (h) Van. For the three last samples, it wasn't necessary to perform acid treatment because phyllosilicate phases were not present in these samples.

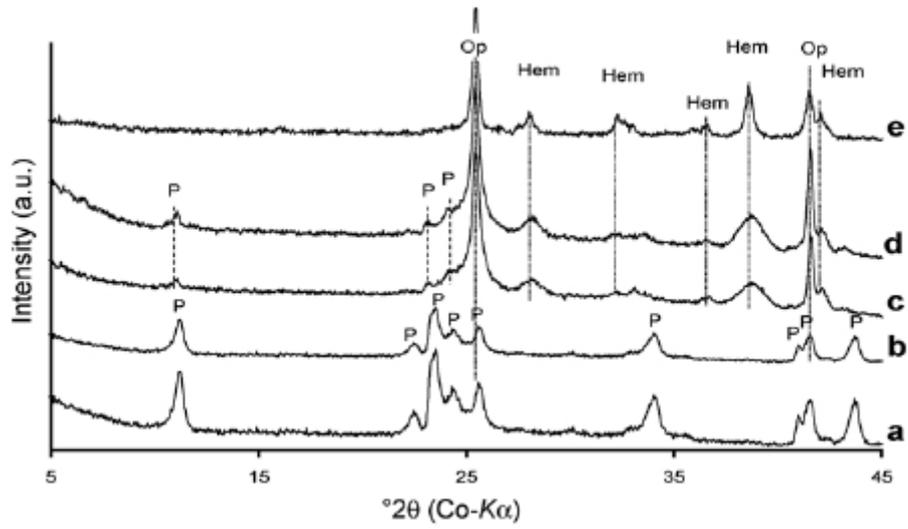

**Figure 5:** XRD patterns of synthetic pyrophyllites. (a) PAl-1, (b) PFe-1, (c) PFe-2, (d) PFe-3. Pyrophyllite reflections are labeled P. The presence of opal and hematite annex phases are labeled Op and Hem respectively.

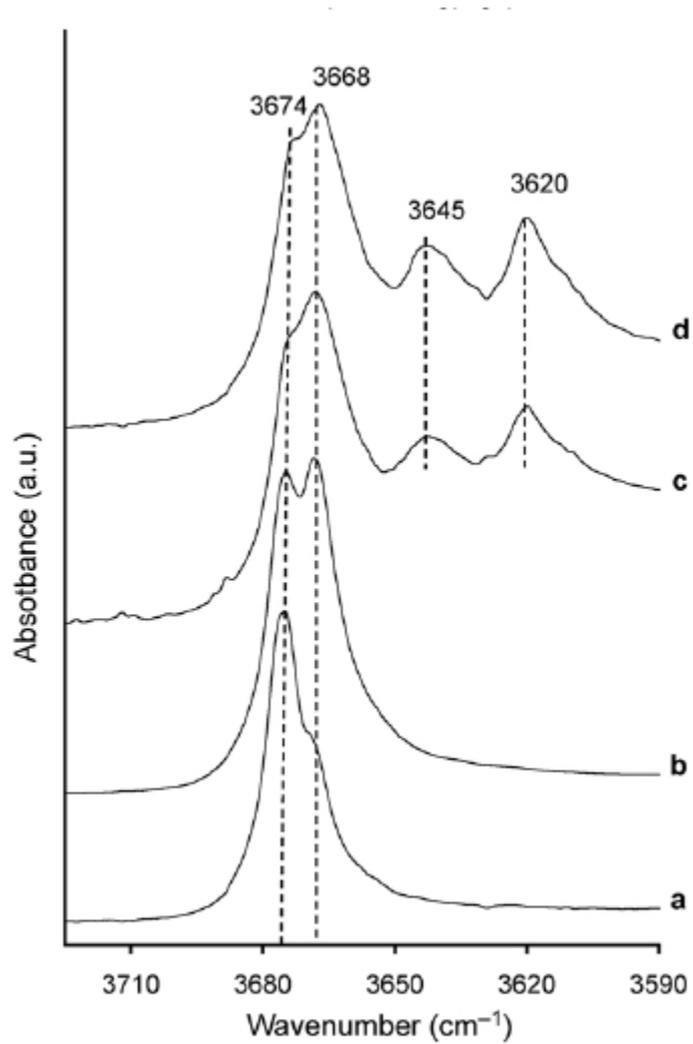

**Figure 6:** Infrared spectra of synthetic samples containing pyrophyllite phase in the OH stretching zone. (a) PAl-1, (b) PAl-2, (c) PFe-1, (d) PFe-2.

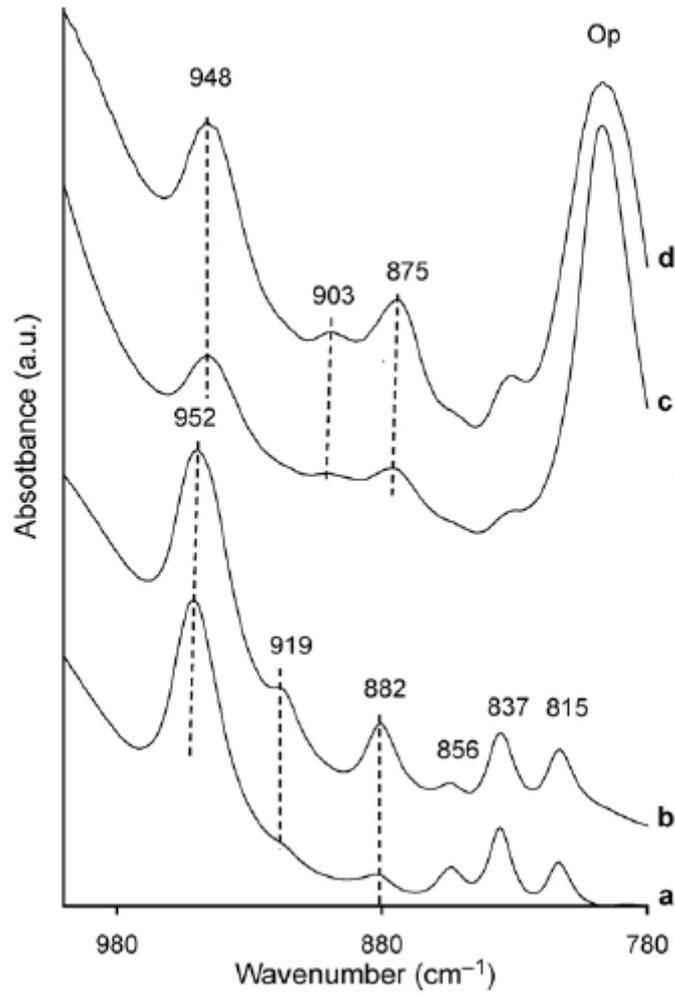

**Figure 7:** Infrared spectra of synthetic pyrophyllites in the OH bending zone. (a) PAl-1, (b) PAl-2, (c) PFe-1, (d) PFe-2.

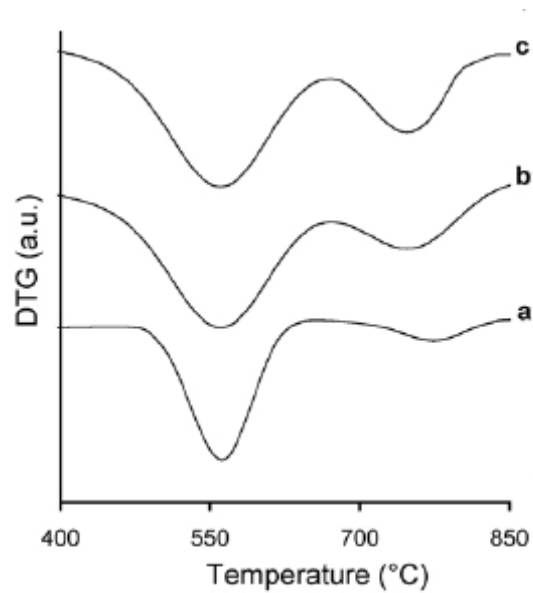

**Figure 8:** DTG curves of pyrophyllites in the 400-850°C temperature range. (a) Guad, (b) PAl-1 and (c) PAl-2 samples.

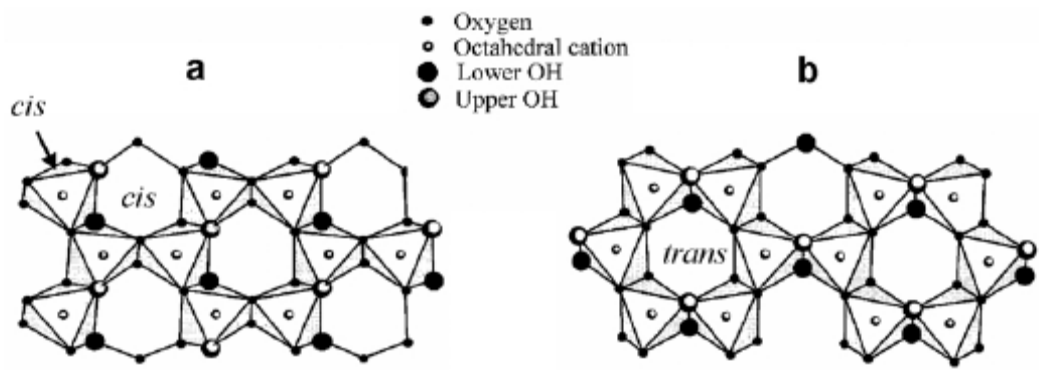

**Figure 9:** A fragment of the octahedral sheet of a 2:1 layer showing the local disposition of the octahedral sites around the OH group. $M_1$ corresponds to the *trans*-site. $M_2$ and $M_{2'}$ are the two *cis*-positions. The OH group are represented by full black circles.

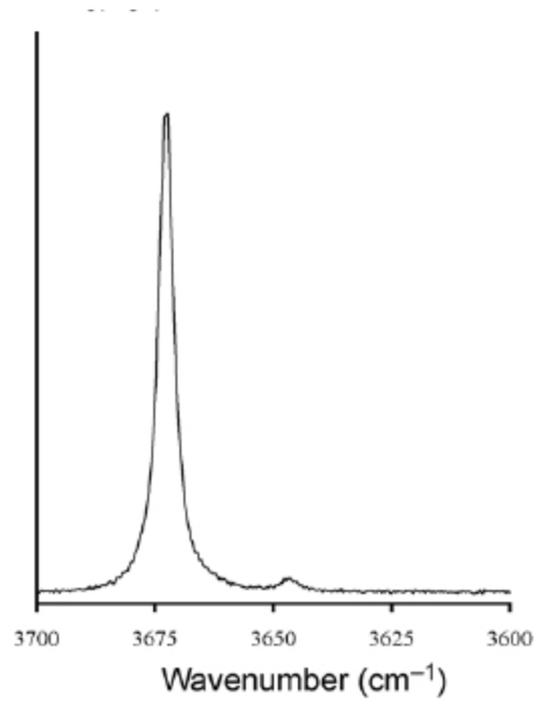

**Figure 10**: Raman spectra of VAN pyrophyllite sample in the OH stretching zone.

**Table 1:** natural pyrophyllite samples examined

| Name | Source | Annex phases associated* | Structure* |
|---|---|---|---|
| Rob 48 | Robbins, N. Carolina, BRGM collection | Mica traces | Monoclinic |
| Rob 49 | Robbins, N. Carolina, BRGM collection | Muscovite and Quartz | Essentially monoclinic |
| Nep | Nepal, Lantenois personal collection | - | Monoclinic |
| Van | Vanoise, France, ENS Paris collection, | - | Monoclinic |
| Sued | Sweden, Ref : 6534, Paris Museum collection | Kaolinite and Muscovite | Monoclinic |
| SN | St. Niclaus valley, Zermatt, Swizerland. | - | Monoclinic |
| Bin | Besson personal collection** | Quartz | Monoclinic |
| Guad | Guadeloupe (france), BRGM collection | Quartz | triclinic |

Notes : * determined by X-ray diffraction, ** Besson et al. [6, 7]

**Table 2:** Chemical and experimental data of synthetic pyrophyllites

| Samples | PAl-1 | PAl-2 | PFe-1 | PFe-2 | PFe-3 |
|---|---|---|---|---|---|
| Chemical composition of beginning gel phases[1]: | | | | | |
| $SiO_2$ | 63.8 | 66.2 | 64.3 | 57.3 | 43.4 |
| $Al_2O_3$ | 36.2 | 33.8 | 28.9 | 5.0 | 0.0 |
| $Fe_2O_3$ | 0.0 | 0.0 | 6.6 | 36.6 | 56.4 |
| $Na_2O$ | 0.0 | 0.0 | 0.1 | 0.1 | 0.2 |
| Experimental conditions of hydrothermal synthesis: | | | | | |
| Temperature (°C) | 475 | 430 | 475 | 475 | 475 |
| Pressure (kbar) | 2 | 0.5 | 2 | 2 | 2 |
| Run time (Days) | 15 | 29 | 15 | 15 | 15 |

Note: [1] Determined by ICP measurements

**Table 3:** Chemical formulae of pyrophyllites

| Samples | Rob48 | Rob49 | Rob49$^T$ | Sued | Sued$^T$ | SN | SN$^T$ | Bin | Guad | Nep | Van | PAl-1 | PAl-2 |
|---|---|---|---|---|---|---|---|---|---|---|---|---|---|
| Tetrahedral occupancy | | | | | | | | | | | | | |
| Si | 3.98 | 3.72 | 3.99 | 3.87 | 3.99 | 4.00 | 4.00 | 4.00 | 3.98 | 3.98 | 4.00 | 3.85 | 4.00 |
| Al | 0.02 | 0.28 | 0.01 | 0.13 | 0.01 | 0.00 | 0.00 | 0.00 | 0.02 | 0.02 | 0.00 | 0.15 | 0.00 |
| Octahedral occupancy | | | | | | | | | | | | | |
| Al | 2.00 | 1.96 | 1.99 | 2.20 | 1.97 | 1.97 | 1.97 | 1.97 | 2.00 | 1.98 | 1.97 | 2.00 | 1.98 |
| Fe | 0.00 | 0.00 | 0.00 | 0.03 | 0.03 | 0.03 | 0.03 | 0.03 | 0.00 | 0.02 | 0.03 | 0.00 | 0.00 |
| Interlayer cations | | | | | | | | | | | | | |
| K | 0.01 | 0.36 | 0.00 | 0.07 | 0.01 | 0.00 | 0.00 | 0.00 | 0.01 | 0.02 | 0.00 | 0.00 | 0.00 |
| Na | 0.01 | 0.04 | 0.00 | 0.01 | 0.00 | 0.00 | 0.00 | 0.00 | 0.00 | 0.01 | 0.00 | 0.00 | 0.00 |

**Note:** all chemical formulae were calculated using electron microprobe analyses. $^T$ pyrophyllite samples after acid treatment.